\documentclass[preprint]{revtex4}
\usepackage[dvipdfmx]{graphicx}
\usepackage{ulem}
\usepackage{comment}
\begin{document}

\title{Quantum statistical effects in warm nuclear matter with light and heavy clusters}

\author{Shun Furusawa}
\email{furusawa@rs.tus.ac.jp}
\affiliation{Department of Physics, Tokyo University of Science, Shinjuku, Tokyo, 162-8601, Japan}
\affiliation{Interdisciplinary Theoretical and Mathematical Sciences Program (iTHEMS), RIKEN, Wako, Saitama 351-0198, Japan}
\author{Igor Mishustin}
\affiliation{Frankfurt Institute for Advanced Studies, J.W. Goethe University, 60438 Frankfurt am Main, Germany}
\affiliation{National Research Center Kurchatov Institute, Moscow 123182, Russia}

\date{\today}

\begin{abstract}
We have investigated the compositions of hot and dense nuclear matter
with the focus on the quantum-statistical effects for  light clusters. 
Our main observation is that
the formation of heavy nuclei in stellar matter  leads to the reduction of  the number densities of light clusters.
As a result, the quantum statistical  effects such as  Bose-Einstein condensation
of deuterons and $\alpha$-particles 
are suppressed. The deviations in number densities of light clusters between  Boltzmann and  quantum statistics at sub-nuclear densities and temperatures 1-3 MeV are $\sim$0.2$\%$ at most.
The condensation of $\alpha$-particles in iso-symmetric nuclear matter is predicted 
only under the assumptions that it is composed  only of nucleons and light clusters
and no heavy nuclei are present.
We also found that the Coulomb screening effects hardly affect  the critical baryon densities for alpha condensation, 
 although mass and chemical potential of  $\alpha$-particles are modified.
\end{abstract}

\pacs{}

\maketitle

\section{Introduction \label{intro}}
Hot and dense stellar matter appears in such astrophysical phenomena  as core-collapse supernovae and neutron-star mergers.
The nuclear matter at sub-nuclear densities is made of  mixture of unbound nucleons, light clusters, and heavy nuclei
in chemical and thermal equilibrium.
 Nuclear equation of state (EOS)  determines  number densities of all these nuclear 
species  and thermodynamical quantities as functions of temperature, baryon density, and charge fraction. 
Dynamical evolution and observable signatures 
of such compact-star  phenomena are very sensitive to the EOS.
In addition, weak interactions of the nuclear species play an important role, especially in core-collapse supernovae (see e.g. \cite{furusawa17b}).
The roles of light clusters in core-collapse supernovae have not been completely clarified,
while their  potential influence on  the dynamics,
 neutrino spectrum, and nucleosynthesis has been pointed out by many researchers 
\cite{haxton88,oconnor07,ohnishi07,sumiyoshi08,arcones08,langanke08,barnea08,furusawa13b,nasu15,fischer16,nagakura19a}.

Modern EOSs for clustered nuclear matter at sub-nuclear densities for astrophysical simulations 
are usually obtained by using  statistical models, e.g. \cite{botvina10, hempel10,furusawa11,buyukcizmeci14,furusawa13a,furusawa17a,furusawa17d, furusawa18b, schneider17, pais19, furusawa19a}.
Recently we have developed a self-consistent  statistical approach where individual nuclear ground state properties and the  
 full-ensemble distributions of nuclei are calculated consistently  by minimizing free energy density \cite{furusawa17c,furusawa18a}. 
In most  statistical models, thermodynamical quantities of light clusters are calculated with Boltzmann statistics.
This assumption is actually valid at high temperatures and low densities.
It is a priori unclear whether this assumption is valid for conditions encountered in supernovae  and neutron star mergers.
Certainly, at low temperatures and at high densities,
they should be regarded as Fermi or Bose particles.

The purpose of this work is to investigate such quantum statistical  effects for light clusters
in stellar environments  with realistic nuclear ensembles. 
The possibility of Bose-Einstein condensation of light clusters in nuclear matter
was earlier studied  in refs. 
\cite{misicu17, sedrakian17,wu17,satarov17,satarov19,zhang19}. These previous studies, however, did not
consider  the full nuclear ensemble including light and heavy nuclei. 
In the present study, we generalize the classical models of light clusters to the quantum statistical description
 in the framework of our statistical approach including full ensemble of nuclei.

We formulate the modified statistical model in Sec.~\ref{sec:model} with a focus on the quantum statistical effects of light clusters.
The results of calculations with the full statistical ensemble including light and heavy nuclei at  some typical astrophysical conditions are discussed in Sec.~\ref{sec:res1}.
For comparison, the clustered nuclear matter consisting only of nucleons and light clusters is considered in Sec~\ref{sec:res2} to demonstrate a crucial role by heavy nuclei. 
Conclusions are  presented in Sec.~\ref{sec:conc}.

\section{statistical model of clustered nuclear matter with quantum  statistics
 \label{sec:model}}
To describe multi-component nuclear matter with light and heavy clusters,
 we use  the same models as in our previous work 
\cite{furusawa18a}.
The free energy density of uniformly-distributed nucleons is evaluated 
with Skyrme type interactions \cite{oyamatsu03,oyamatsu07} and  finite temperature expressions for kinetic energies.
The free energies of heavy nuclei are given by the Boltzmann gases with excluded volume effects and with the mass free energies that are based on the  compressible liquid drop model.  
In the previous works, Maxwell-Boltzmann statistics was employed to evaluate the free energies of both light and heavy nuclei.
In the present work, the Bose-Einstein and Fermi-Dirac statistics are applied to the light clusters with odd and even mass numbers, respectively. 

The total free energy density of the system is given as
\begin{eqnarray}
f = f_{np}  +  \sum_j  f_{lj} + \sum_i  n_i (F^t_{i} + M_{i}) ,
\label{total}
\end{eqnarray}
where $f_{np}$  is the free energy densities of dripped nucleons and $f_{lj}$ are the free energies of light nuclei $j$ with the atomic number $Z_j \leq 5$.
The index $i$ runs over all nuclear species of heavy nuclei with $6 \leq Z_i \leq 1000$.
In Eq. (\ref{total}),  $n_i$, $F_i^t$, and $M_i$ are the number densities, translational free energies, and mass free energies of nuclei $i$.
The calculations of $f_{np}$, $n_i$, $F^t_{i}$, and $M_{i}$ are the same as in ref. \cite{furusawa18a},
where one can also find all the detail.

\renewcommand\thefootnote{\alph{footnote}}
In this work, the light clusters  with even/odd mass numbers are assumed to obey the Bose/Fermi statistics
$^{\rm \footnotemark[1]}$.
 \footnotetext[1]{Below the units $\hbar=c=1$ are used.} 
\begin{eqnarray}
f_{lj} &=& n_j \mu_j - p_j , \\
\label{lightclusters}
n_j &=& g_{j}  \lambda_{j}^{-3} \frac{2F_{1/2} (\eta_j) }{\sqrt{\pi}}   \label{eq_nd} , \\
p_j &=& g_{j}  \lambda_{j}^{-5} \frac{F_{3/2}(\eta_j)}{6\sqrt{\pi} M_j}. 
\end{eqnarray}
Here  $\mu_j$ and $p_j$ are chemical potential and pressure of light cluster $j$, $g_j$ is the ground-state degeneracy factor,  $\eta_j = (\mu_j-M_j)/T$, $T$ is temperature,  $\lambda_{j}=\sqrt{2 \pi /(M_j T)}$,  and 
\begin{eqnarray}
F_k(\eta) = \int_0^{\infty} z^k \left[ {\rm exp}(z-\eta) \pm  1 \right]^{-1}  dz 
\end{eqnarray}
are Fermi $(+)$ and Bose $(-)$ integrals. 
In the low-density and high-temperature limit ($n_j \ll \lambda_j^{-3}$), 
the free energy  is equal to that of Boltzmann gas as  was assumed  in our  previous works \cite{furusawa17c,furusawa18a}.   
The masses of light clusters are assumed to be experimental masses, 
$M_j^{ex} $, with Coulomb energy shifts, $\Delta E_j^C $:
 \begin{eqnarray}
M_j &=&M_j^{ex} +\Delta E_j^C , \\
\Delta E_j^C(n'_p,n_e)&=& \frac{3}{5} \left(\frac{3}{4 \pi} \right)^{-1/3}  e^2 n_{0}^2 {V_{Nj}}^{5/3}
 \left[ \left(\frac{Z_j - n'_p V_{Nj}}{A_j}\right)^2   (1-\frac{3}{2}u_j^{1/3}+\frac{1}{2}u_j) -  \left(\frac{Z_j}{A_j}\right)^2  \right] ,  \label{eqcl}
\end{eqnarray}  
where $A_j=Z_j+N_j$,  $V_{Nj}=A_j/n_0$, $u_j=(n_e-n'_p)/(Z_j/V_{Nj}-n'_p)$,  $n_e$ is electron density,  and $n'_p$ is local proton density in the vapor,
see details in refs. \cite{furusawa11,furusawa17c}. The same formula for the Coulomb energy is also applied for heavy nuclei in stellar environments.

\begin{figure}
\begin{center}
\includegraphics[width=9 cm]{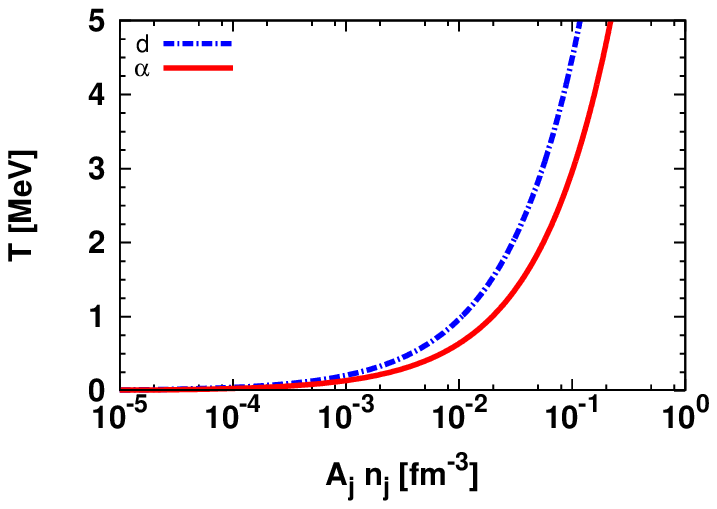}
\end{center}
\caption{Critical temperatures below which Bose-Einstein condensation are realized as a function of the baryon density (mass number times number density) for deuterons ($A_j=2$, blue dashed-dotted lines) and $\alpha$-particles ($A_j=4$, red thick solid lines). 
}
\label{fig_cd}
\end{figure}

Bose particles form a condensate under the condition, $n_j>n_{jC}$, where the critical density, $n_{jC}$ is given by Eq.~(\ref{eq_nd}) after substitution $\eta_j=0$ (or $M_j=\mu_j$).
The critical temperature below which Bose particles form a condensate can be expressed as
 \begin{eqnarray}
T_{jC}=\frac{2 \pi }{M_j} \left(\frac{n_{j}}{g_j \zeta(3/2)}  \right)^{2/3} , \label{eq_cd}
\end{eqnarray}  
where $\zeta(x)$ is zeta function, $\zeta(3/2)=2.612$.
Figure~\ref{fig_cd} displays the critical lines in ($A_j n_{j}$, $T$) plane. 
 As one can see,  at high densities and low temperatures, the condensation of $\alpha$-particles  can be , \d. 
Here, we ignore the existence of electrons and dripped protons, which affect nuclear Coulomb energies.
The critical baryon number densities, $A_j n_{jC}$, of deuterons are lower than those of $\alpha$-particles for the same temperature. 
The condensation of deuterons, however, is less likely to be reached, because their number density and binding energy are considerably smaller than those of $\alpha$-particles.

\section{warm clustered matter with realistic nuclear ensemble \label{sec:res1}}
\begin{figure}
\begin{center}
\includegraphics[width=8.1cm]{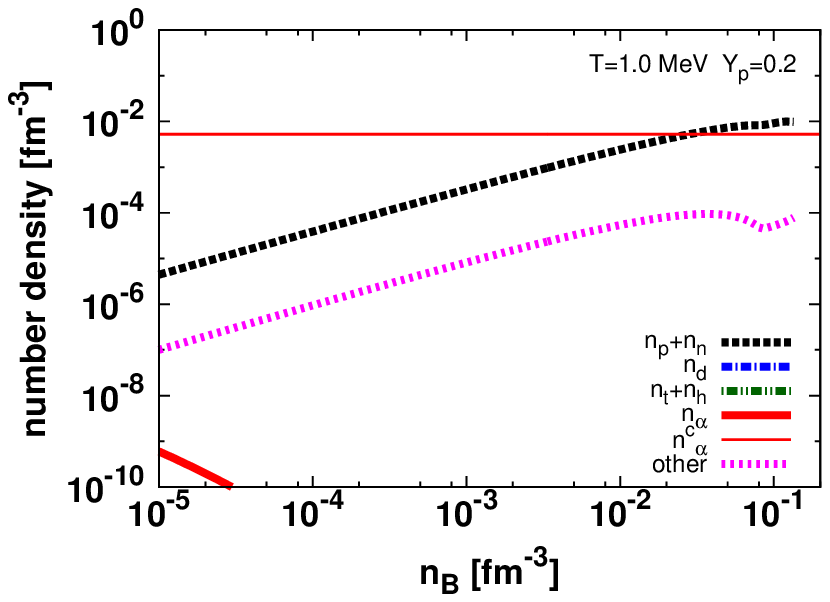}
\includegraphics[width=8.1cm]{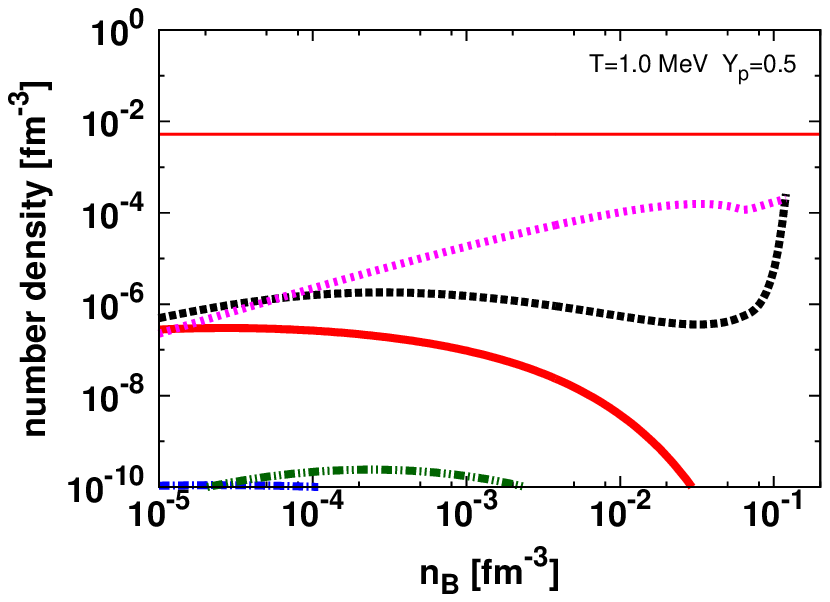}
\includegraphics[width=8.1cm]{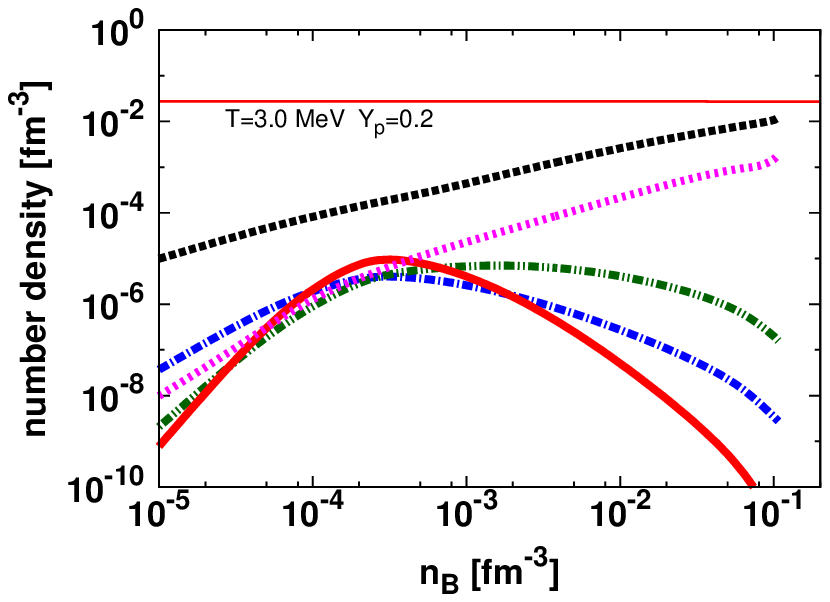}
\includegraphics[width=8.1cm]{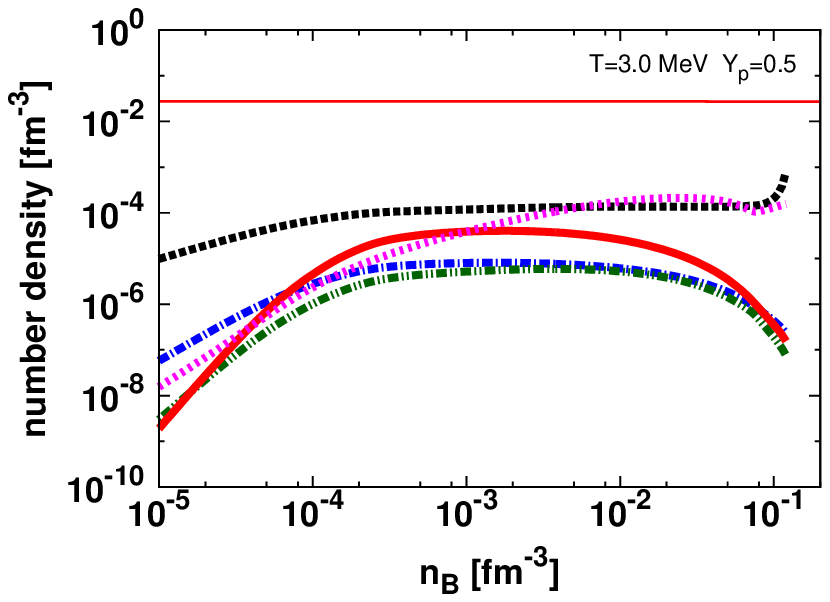}
\end{center}
\caption{
Number densities of dripped protons and neutrons (black dashed lines), deuterons (blue dashed dotted lines), tritons and helions (green dashed double-dotted lines), $\alpha$-particles (red thick solid lines), and  the other nuclei  ($Z_i>2$ or $N_i>2$,  magenta dotted lines) as functions of  $n_B$
at  $T=$1.0 MeV (top row) and  3.0~MeV (bottom row)
and $Y_p=$ 0.2 (left column) and 0.5 (right column).
Red thin solid lines display the critical number density above which $\alpha$-particles should form a condensate.
}
\label{fig_nd}
\end{figure}

\begin{figure}
\begin{center}
\includegraphics[width=8.1cm]{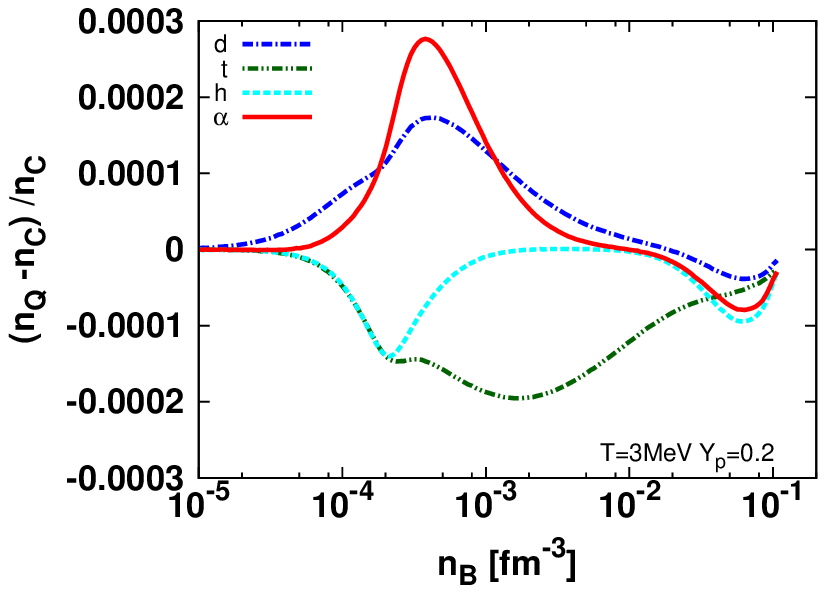}
\includegraphics[width=8.1cm]{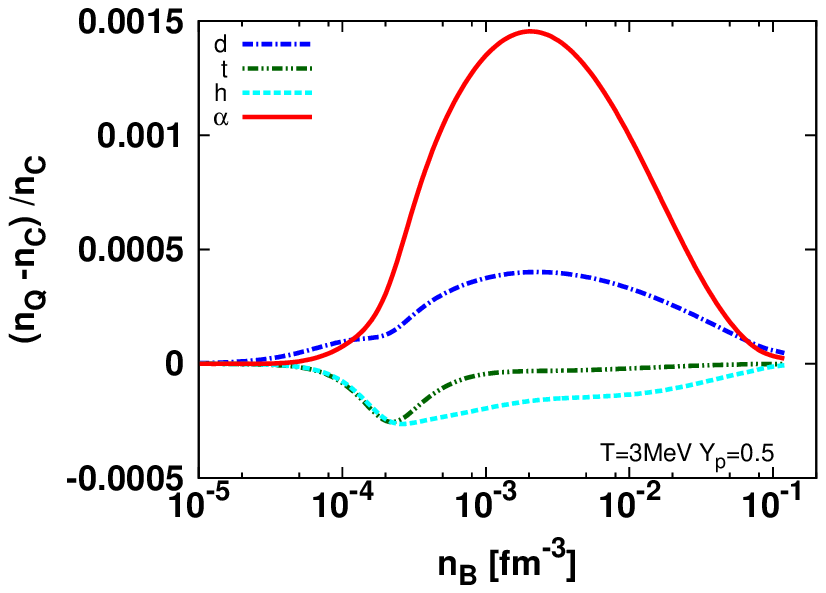}
\end{center}
\caption{Relative change  of number densities of  deuterons (blue dashed dotted lines), tritons (green dashed double-dotted lines), helions (cyan dashed lines),  and $\alpha$-particles (red thick solid lines) calculated in quantum statistics to those for classical (Boltzmann) statistics at  $T=$~3.0~MeV
and  $Y_p=$ 0.2 (left panel) and 0.5 (right panel).
}
\label{fig_rat}
\end{figure}

\begin{figure}
\begin{center}
\includegraphics[width=8.1cm]{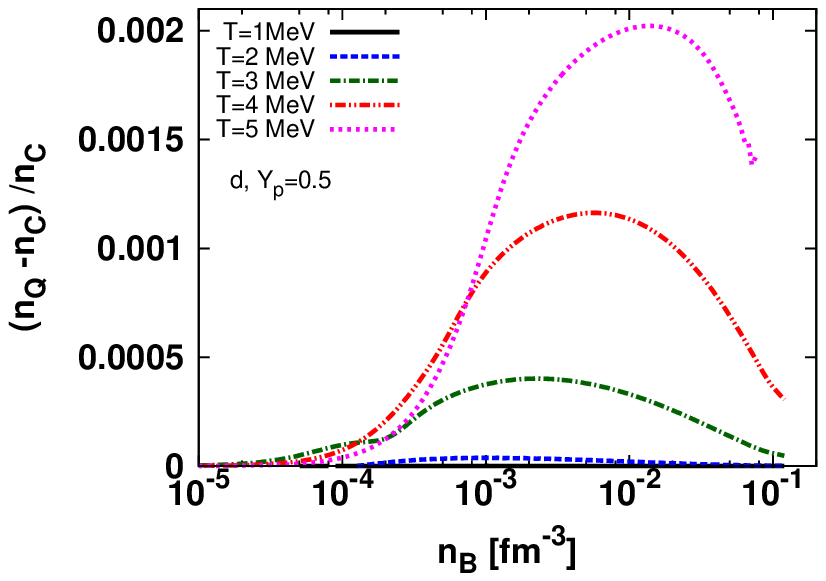}
\includegraphics[width=8.1cm]{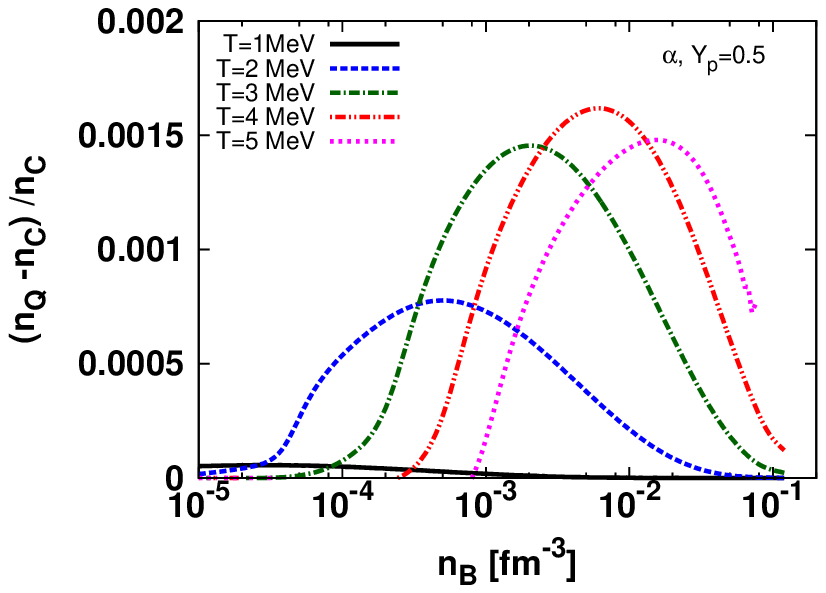}
\end{center}
\caption{Relative change of number densities of deuterons (left panel) and  $\alpha$-particles (right panel) calculated in quantum statistics 
to those for classical (Boltzmann) statistics
 at  $Y_p=$~0.5 and T=1.0 MeV (black solid lines), 2.0 MeV (blue dashed lines), 3.0 MeV (dashed dotted lines), 4.0 MeV (dashed double-dotted lines), and 5.0 MeV (dotted lines).
}
\label{fig_rat2}
\end{figure}

 The number densities of dripped-protons and -neutrons, deuterons (d), tritons (t), helions  (h), $\alpha$-particles ($\alpha$),  and other clusters are given in Fig.~\ref{fig_nd}. 
We find that Bose-Einstein condensation of light  clusters is unlikely 
 because free nucleons and/or heavy nuclei are dominant components of the nuclear matter under considered conditions.
In other words, they consume almost all nucleons making densities of light bosonic clusters insufficient to create a condensate.
This result means that the inclusion of heavy nuclei is crucial
for realistic description of 
stellar matter.
Figures~\ref{fig_rat} and \ref{fig_rat2} show the relative change
 $(n_j^Q-n_j^C)/n_j^C$ of densities of light clusters calculated with 
Bose or Fermi statistics, $n_j^Q$, as compared to those in Boltzmann statistics, $n_j^C$.
We find that in a multicomponent nuclear matter,
there is little difference in the number densities of light clusters between the  quantum  and classical statistics.
The deviations are less than $\sim 0.20\%$. 
In neutron-rich matter, $Y_p=0.2$, and high densities, $n_B \gtrsim 10^{-3}$~fm$^{-3}$, the chemical potentials of neutrons are quite larger than those of protons and, as a result, 
the quantum statistical effects for tritons are stronger than those for $\alpha$ particles.

In some previous works, the contribution of  heavy nuclei was neglected or represented
 by a single nucleus such as Fe$^{56}$ \cite{sedrakian17, wu17} or an optimized nucleus \cite{pais19}.
The single nucleus approximation considerably underestimates the total mass fraction of heavy nuclei and
 overestimates the average mass number of heavy nuclei, even if the mass and proton numbers of the representative nucleus are optimized as shown in ref. \cite{furusawa17c}. 
In the single nucleus approximation, the free energy density is expressed as  $f= f_{np}  +  \sum_j  f_{lj}  + n_{rep} F_{rep}(A_{rep},Z_{rep})$ instead of Eq.~(\ref{total}),
where $n_{rep}$ and $F_{rep}$ are, respectively, the number density and free energy of  a representative nucleus. 
Its mass and proton numbers, $A_{rep}$ and $Z_{rep}$, are optimized by the conditions  ${\partial F_{rep}}/{ \partial A_{rep}}|_{Z_{rep}}=\mu_n$ and ${\partial F_{rep}}/{ \partial Z_{rep}}|_{A_{rep}}=\mu_{p}-\mu_{n}$. In addition, we assume chemical equilibrium among nucleons,  light clusters and the representative nuclei under baryon and charge conservations (see details in ref. \cite{furusawa17c}). Here $\mu_p$ and $\mu_n$ are chemical potentials of protons and neutrons.

\begin{figure}
\begin{center}
\includegraphics[width=8.1cm]{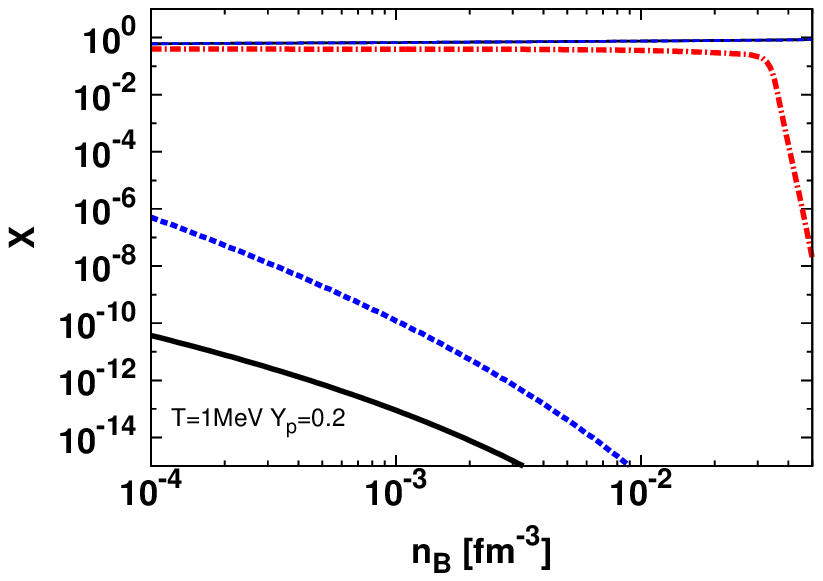}
\includegraphics[width=8.1cm]{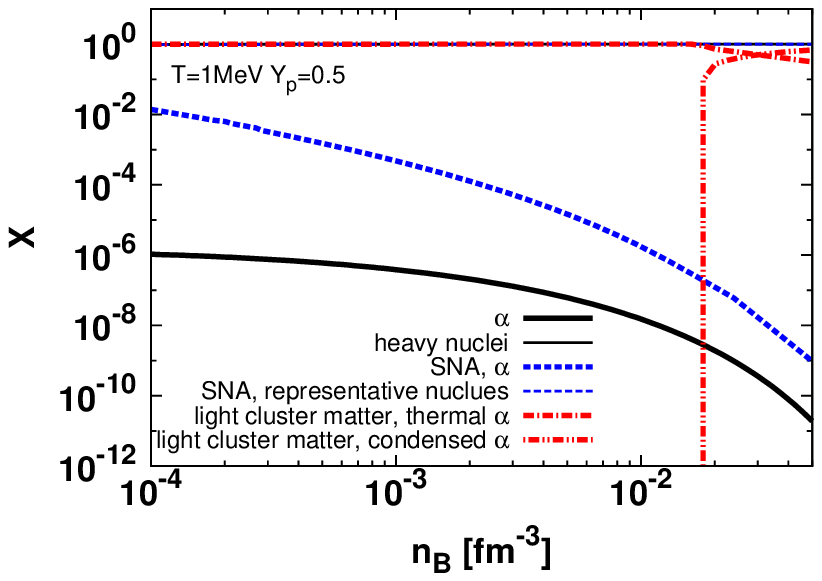}
\includegraphics[width=8.1cm]{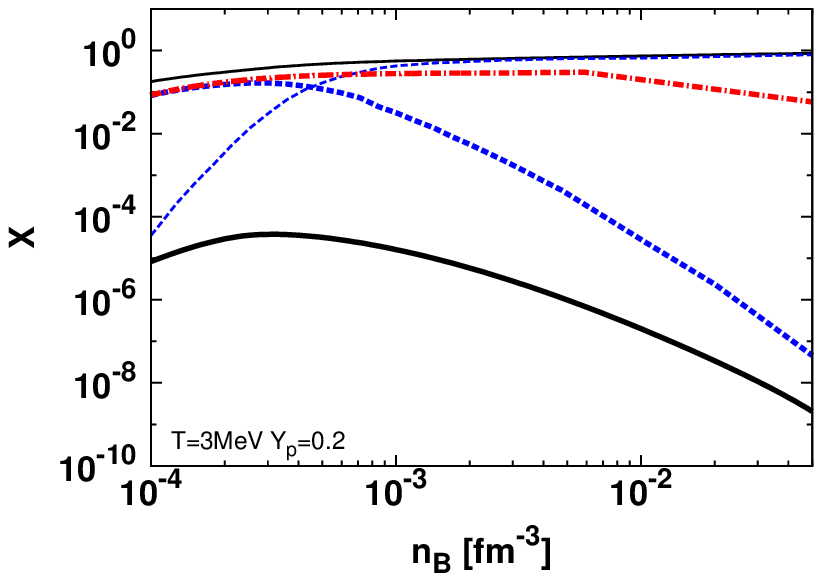}
\includegraphics[width=8.1cm]{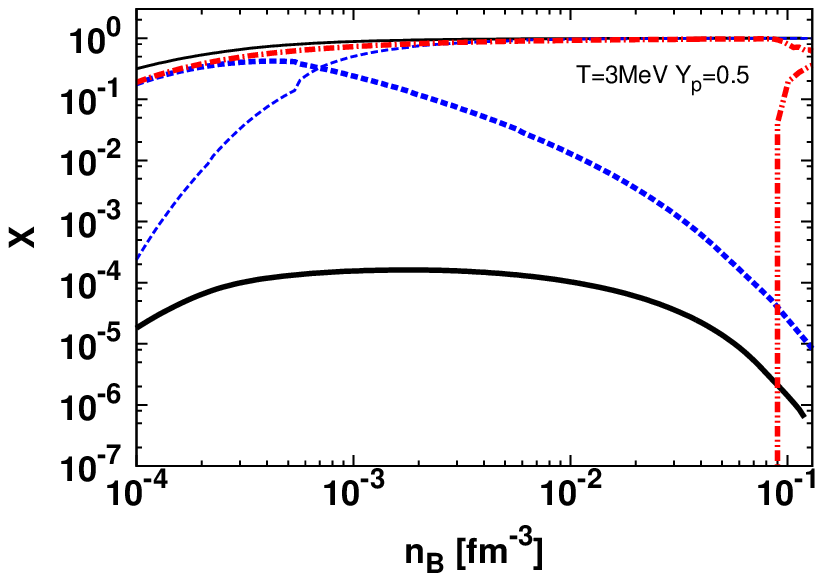}
\end{center}
\caption{Mass fractions of  $\alpha$-particles (thick lines)  and heavy nuclei (thin lines) 
for the model solving full-ensemble of heavy nuclei  (black solid lines) and the model with single nucleus approximation for heavy nuclei ($Z_i>5$) (blue dashed line)
at  $T=$1.0 MeV (top row) and  3.0~MeV (bottom row)
and $Y_p=$ 0.2 (left column) and 0.5 (right column).
The red lines indicate the mass fraction of  thermal $\alpha$-particles (dashed dotted lines) and condensed $\alpha$-particles (dashed double-dotted lines) for the light cluster matter in which we ignore the existence of heavy clusters other than $d$, $t$, $h$, and $\alpha$. }
\label{fig_sna}
\end{figure}

Figure~\ref{fig_sna} shows the mass fractions of $\alpha$-particles for the realistic nuclear ensemble, the single nucleus approximation, and
 the model with no heavy nuclei with $Z>2$ or $N>2$, that is discussed in more detail in the next section. 
As one can see from the figure, the mass fraction of $\alpha$-particles in the more realistic ensemble is considerably smaller than in  the other more restrictive models, due to  the larger population of heavy nuclei.
Indeed,  the mass fractions of heavy nuclei other than representative nucleus  in the single-nucleus  model, and mass fractions
of nuclei other than $d$, $t$, $h$, and $\alpha$  in the light-cluster matter  are strictly zero.
 As a result, the mass fractions of  $\alpha$ -particles in these two models are larger  than in the full-ensemble model. 

\section{warm light-cluster matter with restricted nuclear ensemble \label{sec:res2}}
\begin{figure}
\begin{center}
\includegraphics[width=8.1cm]{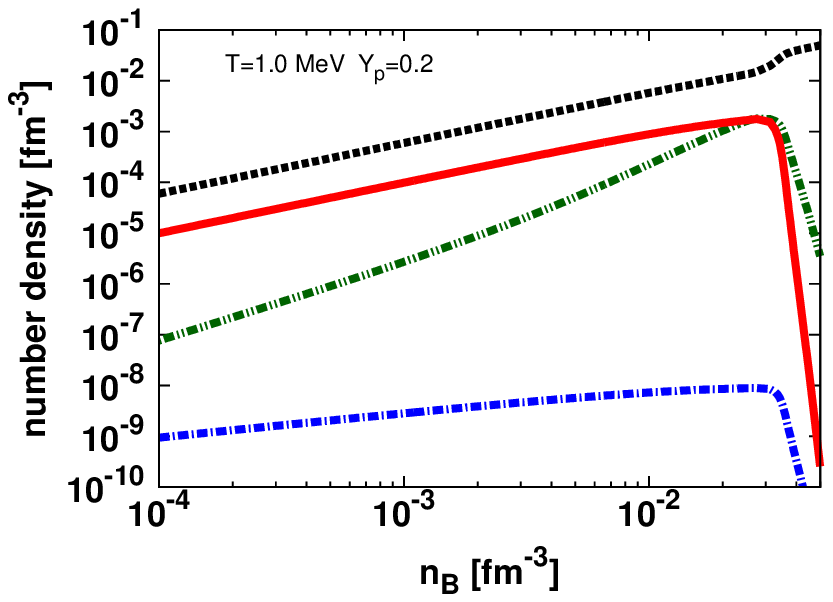}
\includegraphics[width=8.1cm]{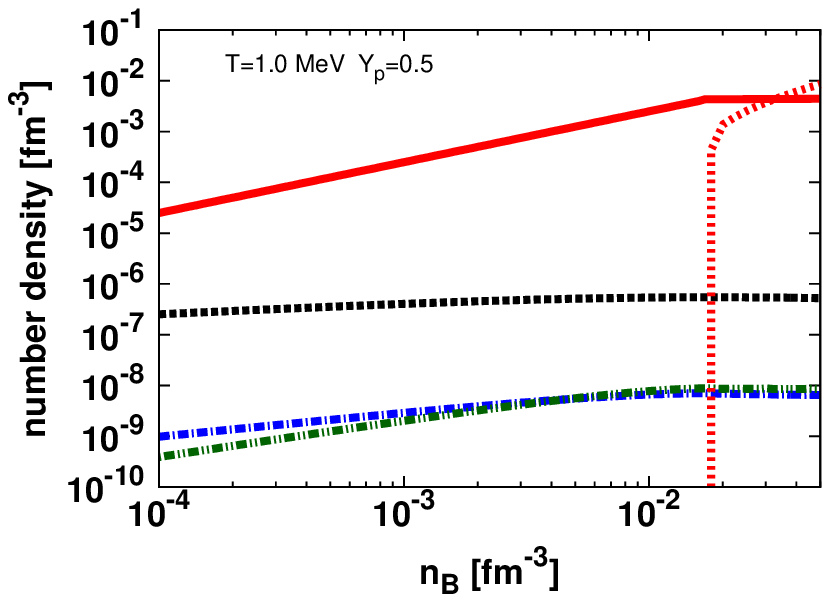}
\includegraphics[width=8.1cm]{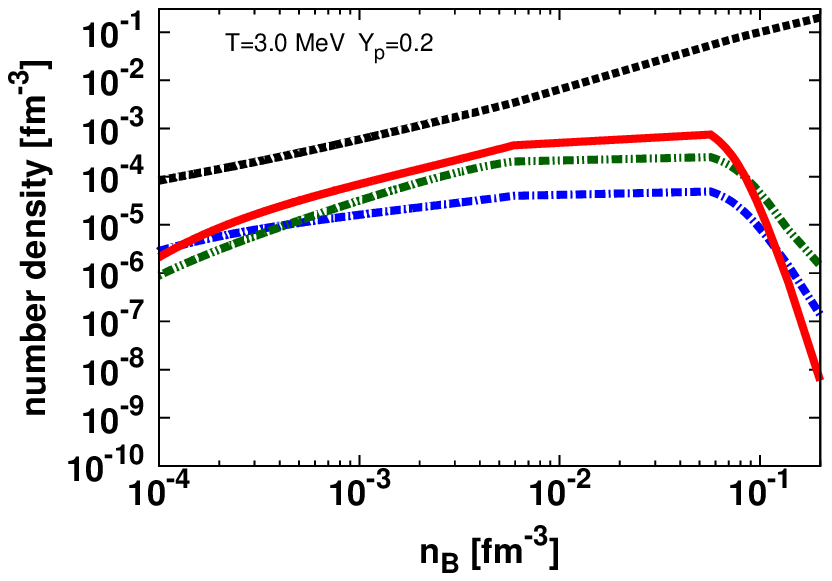}
\includegraphics[width=8.1cm]{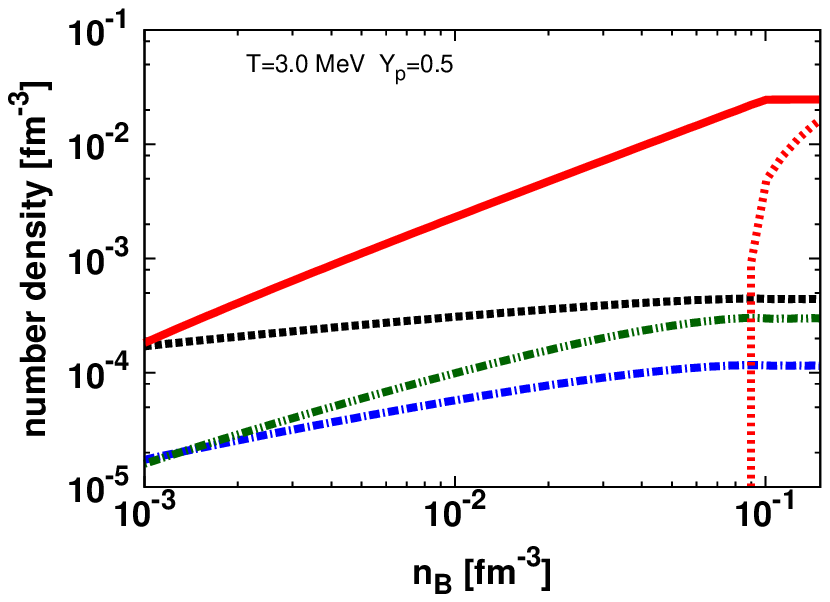}
\end{center}
\caption{Number densities of dripped protons and neutrons (black dashed lines), deuterons (blue dashed dotted lines), tritons and helions  (green dashed double-dotted lines), thermal $\alpha$-particles (with kinetic energy $\epsilon>0$, red thick solid lines), and condensed $\alpha$-particles ($\epsilon=0$, red dotted lines) as functions of  $n_B$
for the light cluster matter
at  $T=$1.0 MeV (top row) and  3.0~MeV (bottom row)
and $Y_p=$ 0.2 (left column) and 0.5 (right column).}
\label{fig_ln}
\end{figure}

In this section, we consider clustered nuclear matter where heavy elements ($Z>2$ or $N>2$) are artificially suppressed, e. g. only $d$, $t$, $h$, and $\alpha$ are allowed
 in the chemical equilibrium.
Then we can  discuss the possibility of alpha condensation under the same assumptions as in ref \cite{zhang19}. 
Figure~\ref{fig_ln} shows  the number densities of nucleons and light clusters
in the model where heavy nuclei are not included. 
As one can see, in this more constrained ensemble,  the $\alpha$-particles may condense at $n_B\sim 0.02$ ($T=1$~MeV) and $0.09$~fm$^{-3}$ ($T=3$~MeV)  in  iso-symmetric matter ($Y_p=0.5$). 

\begin{figure}
\begin{center}
\includegraphics[width=9cm]{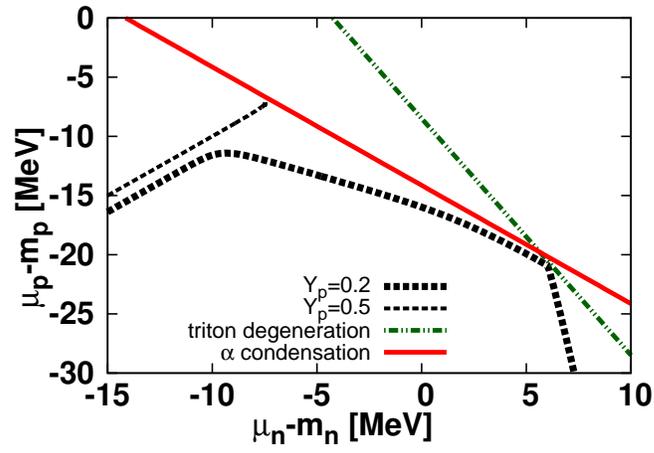}
\end{center}
\caption{Proton chemical potential as a function of neutron chemical potential for the light cluster matter
at  $T=$1.0 MeV  and  $Y_p=$ 0.2 (thick dashed black line) and 0.5 (thin dashed black line) and critical lines at which the the chemical potentials of tritons and $\alpha$-particles reach their rest masses without the Coulomb energy shifts, $\mu_p+2\mu_n=m_t$ (green dashed double-dotted line) and
 $2\mu_p+2\mu_n=m_{\alpha}$ (red solid line). }
\label{fig_ce}
\end{figure}

For neutron-rich matter ($Y_p=0.2$), light clusters are less abundant, and  $\alpha$-particles do not condense
even  if the existence of heavy clusters is ignored. 
This is because  proton chemical potentials are much lower than those for $Y_p=0.5$ especially at high densities
as also shown in Fig.~\ref{fig_rat}. 
Figure~\ref{fig_ce} displays the chemical potentials of nucleons for $Y_p=$0.2 and 0.5 and the critical lines at which the chemical potentials of tritons and $\alpha$-particles are equal to  their rest masses without Coulomb energy shifts: $\mu_p+2\mu_n=m_t$ and  $2\mu_p+2\mu_n=m_{\alpha}$. 
Around  and above the lines,  the differences in number densities between classical and quantum statistics become large for each light cluster.  
For $Y_p=0.5$,  $\alpha$-particles start to condense around the line,
while for $Y_p=0.2$,
the quantum statistical effects for tritons become strong, $\mu_p+2\mu_n \sim m_t$, at lower density than the critical densities for $\alpha$-condensate.

\begin{figure}
\begin{center}
\includegraphics[width=8.1cm]{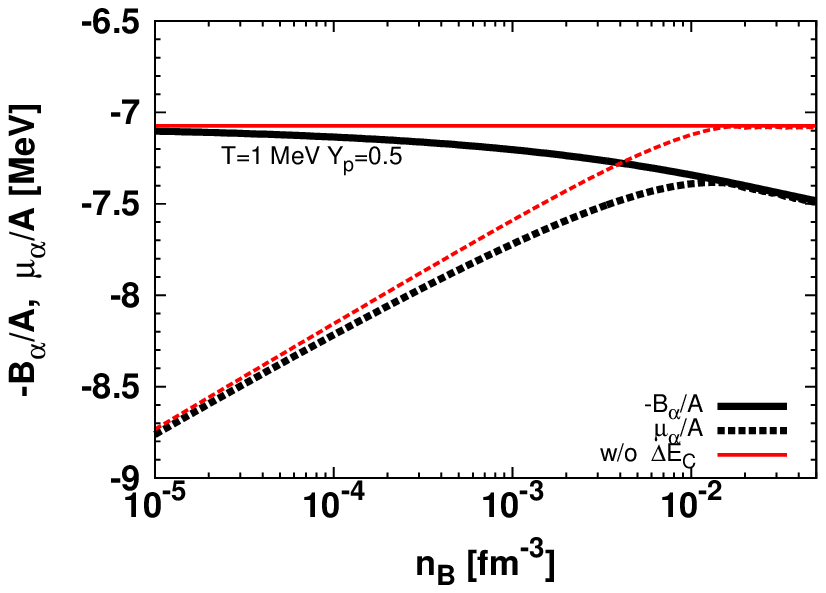}
\includegraphics[width=8.1cm]{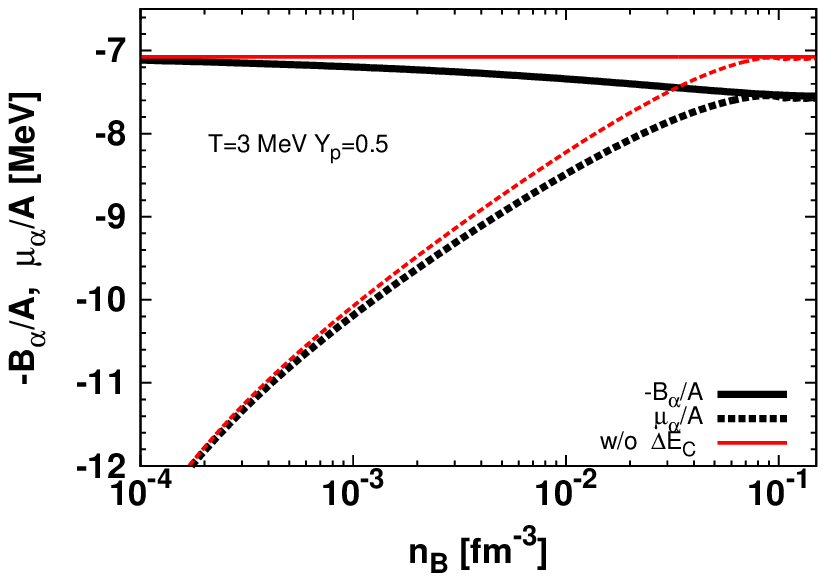}
\end{center}
\caption{Chemical potential of $\alpha$-particles (with respect to its rest mass) per baryon (dashed lines) and  the negative value of its binding energy per baryon (solid lines) as functions of  $n_B$
for the light cluster matter
at  $T=$1.0 MeV (left panel) and  3.0~MeV (right panel)
and $Y_p=$ 0.5.
The red thin  lines display those without the Coulomb energy shifts: $\Delta E_j^C=0$.
}
\label{fig_ch}
\end{figure}

To investigate the impact of Coulomb energy,
we also present  the results obtained without $\Delta E_j^C$ shift. 
Figure~\ref{fig_ch} gives  the chemical potential of $\alpha$-particle relative to its rest mass, $\mu'_\alpha=2 (\mu_n+\mu_p- m_n - m_p)$, per baryon, and
 the negative value of its binding energy, $-B_{\alpha}=M_{\alpha}-2 (m_n + m_p)$, per baryon.
Obviously, the condensation starts when the chemical potential becomes equal to the mass. 
We find that the Coulomb energy shifts actually reduce both the chemical potential and effective mass of $\alpha$-particles.  
Therefore, 
the critical densities at which $\alpha$-particles may condense  depend very weakly  on  the electron distribution.

As shown in  ref. \cite{furusawa18a}, the ground state minimum in the bulk free energy disappears at temperatures around 14 MeV. 
At this point, heavy nuclei disappear from the nuclear ensemble and
 the abundance of $\alpha$-particles may increase significantly. 
However these temperatures are already too high for light-cluster condensation.

\section{Concluding  remarks \label{sec:conc}}
We have investigated the nuclear compositions of hot and dense stellar matter
within  the self-consistent model including full nuclear ensemble. 
In contrast to previous works, here 
the quantum statistics is used for bosonic clusters, while classical statistics is employed for heavier nuclei. 
We find that the presence of heavy nuclei
leads to the reduction of
 number densities of light clusters and, as a result, the condensation of bosonic clusters 
dose not happen. 
We have demonstrated that the differences in number densities of light clusters  between quantum and classical statistics are small,  at most 0.20~$\%$. 
On the other hand,
if we assumed that the iso-symmetric nuclear matter is composed only of nucleons and light clusters, as in some previous works \cite{satarov19,zhang19},  $\alpha$-particles reach the threshold of  Bose-Einstein condensation at rather high densities, $n_B\gtrsim$ 0.01 fm$^{-3}$ for $T=$1--3 MeV. 
%
Such light clusters would be difficult to expect in such dense stellar matter,
where  heavy nuclei are already formed. 
For warm neutron-rich nuclear matter  ($T=$1-3~MeV and $Y_p=0.2$), chemical potentials  of protons and alpha particles are too small for the $\alpha$-condensation, even when the heavy nuclei are not formed.

In this work, the interactions between $\alpha$-particles and other nuclear species are not considered explicitly,
except of excluded volume corrections. 
The strong interaction effects are to some extent also  taken into account by Skyrme-like terms in the energy-density function used for dripped nucleons and bulk energies of heavy nuclei.
In a more realistic approach, one should 
introduce similar mean fields also for light clusters as it was done in refs. \cite{satarov17,satarov19}.
In ref \cite{satarov20}, it is demonstrated for iso-symmetric nuclear matter that at strong enough $\alpha$-N attractive interaction,
$\alpha$-condensate appears even in the nuclear ground state.
In relation to the present work, this means that $\alpha$-condensate may be present inside of heavy nuclei.
In the future, we are planning to investigate this interesting possibility for isospin-asymmetric nuclear matter too. 

\begin{acknowledgments}
S.F. acknowledges H. Tajima and T. Hatsuda for useful discussion. 
This work was supported by JSPS KAKENHI (Grant Number JP17H06365,19K14723)
and  HPCI Strategic Program of Japanese MEXT  (Project ID: hp170304, 180111).
A part of the numerical calculations was carried out on  PC cluster at Center
for Computational Astrophysics, National Astronomical Observatory of Japan.
I. M. acknowledges fruitful discussions with L. M. Satarov and  financial support from Helmholz International Center for FAIR.
\end{acknowledgments}

\bibliography{reference191218}

\newpage

\end{document}